# Colored Image Encryption and Decryption Using Chaotic Lorenz System and DCT2

Mohammed Alsaedi

College of Computer Science

Taibah University, Medina, KSA

msaiedy@gmail.com, masadi@taibahu.edu.sa

*Abstract*-**In this paper, a scheme for the encryption and decryption of colored images by using the Lorenz system and the discrete cosine transform in two dimensions (DCT2) is proposed. Although chaos is random, it has deterministic features that can be used for encryption; further, the same sequences can be produced at the transmitter and receiver under the same initial conditions. Another property of DCT2 is that the energy is concentrated in some elements of the coefficients. These two properties are used to efficiently encrypt and recover the image at the receiver by using three different keys with three different predefined number of shifts for each instance of key usage. Simulation results and statistical analysis show that the scheme high performance in weakening the correlation between the pixels of the image that resulted from the inverse of highest energy values of DCT2 that form 99.9 % of the energy as well as those of the difference image.**

## I. Introduction

The importance of data security has increased with the significant increase in the number of web applications, particularly in governments, banks, and defense establishments. Owing to the open nature of the Internet, which enables real-time transmission, security became an important issue for researchers. During transmission, data privacy may be compromised by a third party. One of the widely used techniques is encryption using chaotic systems. Chaos is sensitive to initial conditions; therefore, its sequences possess randomness and can be manipulated to possess a large key space that can be used to encrypt data such as speech, images, texts, or videos. The characteristics of chaos are internal randomness, sensitive dependence on initial conditions, and overall stable local instability, which refers to its ability to maintain the instability condition after the system is disturbed [1]. In [1], a discretization scheme based on the Euler method is proposed; in this scheme, the Lorenz chaotic system is initialized and it produces the three-dimensional sequences X, Y, and Z. A set of data of these sequences is compared using the autocorrelation and power spectral density characteristics.

The images and videos sizes have a large amount of information content, and the encryption schemes cannot handle emerging problems—such as compression and noise tolerance—in big data [2]. In [2], a random permutation is used; this step is followed by diffusion via a nonlinear chaotic map, which is generated via Lorenz to diffuse the pixels of the text image. Then, the well-known Gram-Schmidt algorithm is used to produce random orthogonal matrices. The scheme uses a symmetric key at the transmitter and the receiver; further, it is claimed that the scheme has the capability of tolerating channel noise and JPEG compression with higher security on encrypting all the pixels in the spatial domain.

In [3], the encryption scheme consists of two levels of encryption. In the first level, the image features are disguised via selective encryption of DCT (Discrete Cosine Transform) coefficients by using XOR operation and pseudorandom sequence to shuffle all the pixels. In the second level, the image features are disguised using a masking technique via convolution. The decryption process involves estimation based on master's signal at the transmitter derivatives using higher-order differentiators. However, in this study, the proposed scheme uses the Lorenz system and the two-dimensional discrete cosine transform (DCT2) of the image to extract the elements that represent 99.9 % of the image energy and uses these elements in the encryption process.

The Lorenz system is a system of ordinary differential equations. The solutions of these ordinary differential equations are chaotic for certain values of the parameters and initial conditions [4-8, 15]. Equations (1) to (3) represent this system.

$$\frac{dx}{dt} = \sigma(y - x) \qquad (1)$$

$$\frac{dy}{dt} = x(\rho - z) - y \qquad (2)$$

$$\frac{dz}{dt} = x * y - \beta * z \qquad (3)$$

where ρ is the Rayleigh number, σ is the Prandtl number, and β is a parameter. Further, x, y, and z are the system states. This system arises in different fields such as electric circuits, lasers, and DC motors. Under certain conditions for ρ and σ, two steady equilibrium points will appear. These conditions are as follows:

$$\begin{cases} \rho > 1 \\ \rho > \sigma \frac{\sigma + \beta + 3}{\sigma - \beta - 1} \\ \sigma > \beta + 1 \end{cases} \qquad (4)$$

The equilibrium points are:

$$(x, y, z) = \begin{cases} \pm \sqrt{\beta(\rho - 1)} \\ \pm \sqrt{\beta(\rho - 1)} \\ \rho - 1 \end{cases} \qquad (5)$$

Typical values for the parameters are ρ=28, σ=10, and β=8/3. If the initial values of $x_0$, $y_0$, and $z_0$ are 2, 1, and 1.05, respectively, and the time interval is between 0 and 50, with $\varepsilon = 10^{-7}$, the Lorenz system has two attractors, as seen in Fig. 1.

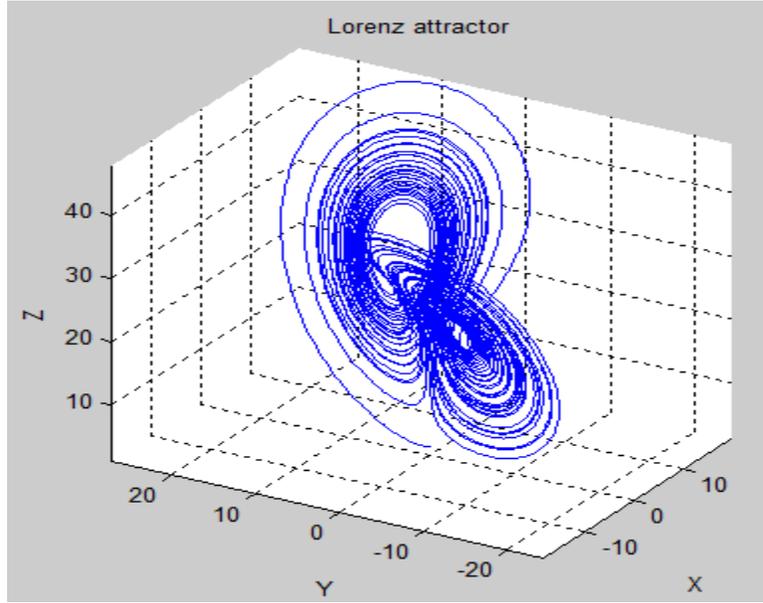

Figure 1. Lorenz attractor

A finite signal f(x, y)—such as audio or an image—can be expressed as a sum of cosine functions using DCT [9-15]. The corresponding mathematical representation is as follows:

$$F(v,u) = \frac{1}{4} C_v C_u \sum_{y=0}^{N-1} \sum_{x=0}^{N-1} f(x,y) \cos\left(\pi v \frac{2y+1}{2N}\right) \cos\left(\pi u \frac{2x+1}{2N}\right) \qquad (6)$$

Fewer cosine functions can be used to approximate a signal. DCT is similar to Discrete Fourier Transform DFT, except that DCT uses real numbers [16-18]. DCT is used for lossy compression because it provides energy compaction, i.e., the energy can be concentrated in a few low-frequency components of DCT that have different amplitudes and fewer frequencies.

The previous Lorenz attractor, shown in Fig. 1, can be represented with fewer components by using DCT. The number of these elements in a 3D system along the X, Y, and Z axes is 208, 286, and 170, respectively. The recovered Lorenz system obtained after performing the inverse discrete cosine transform (IDCT) is shown in Fig. 2.

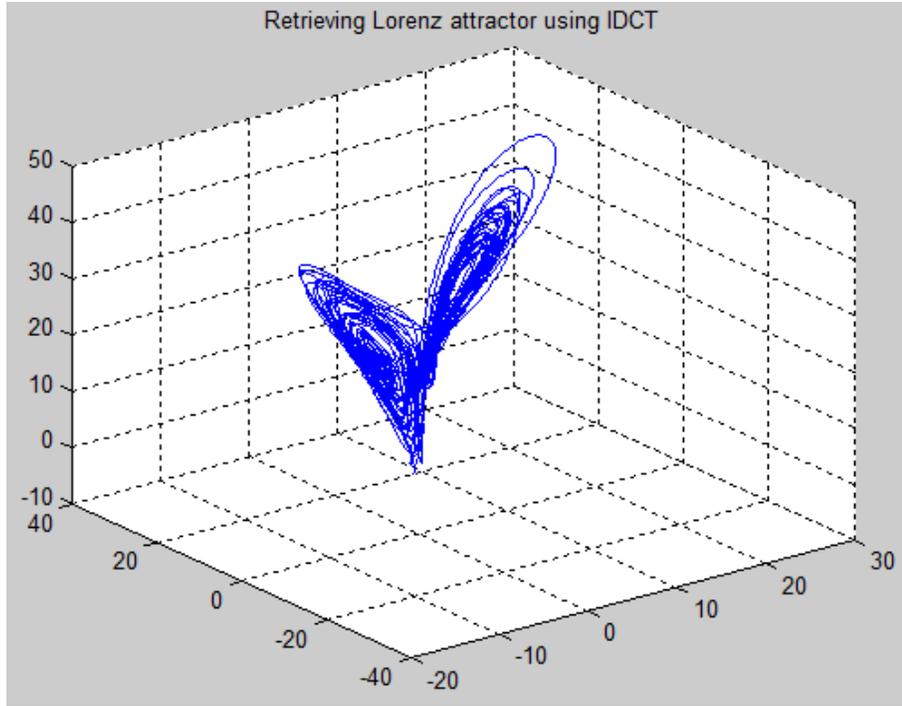

Figure 2. Retrieving Lorenz attractor using IDCT

The Lorenz system and DCT2 can be used in image encryption and decryption; these processes are described in the following sections [19-22]. In Section II, the encryption process in this study is explained; in Section III, the decryption process is described. The simulation results are presented in Section IV. The statistical analysis is discussed in Section V. Finally, the conclusions are stated in Section VI.

## II. ENCRYPTION PROCESS

In the encryption process, three different keys are used with a precision of $10^{-14}$. The first round of encryption process includes a key that contains six characters which are going to be converted to ASCII code bits; then, they are concatenated and shifted cyclically with three different predefined number of shifts; finally, they are rounded with double precision $10^{-14}$. In the encryption process, first, the Lorenz system is initialized with these keys that represent the initial conditions, and the output are three arrays X, Y, and Z, representing the solution of the differential equations. The DCT of these arrays are obtained, and the criteria to retrieve the elements representing 99.9 % of the energy of the signal are applied. Each two arrays are multiplied by each other, thus yielding three matrices of different sizes. These matrices are XY, XZ, and YZ. The next step consists of resizing the matrices to the same size as the input colored image. Further, a convolution of these matrices in two dimensions is performed with a modulus operation of 255. The purpose of this step is to reshuffle the matrices and eliminate equal columns or rows obtained after resizing.

A colored image of size N x N is decomposed into three components. Then, DCT2 is performed; the elements that represent 99.9 % of the energy are retained, whereas the others are ignored. The three 2D arrays obtained from image decomposition are subjected to the inverse discrete cosine transform in two dimensions (IDCT2). The difference between the original decomposed image components and the image obtained as a result of the IDCT2 operation for these components is computed. At this stage, we term the output as the difference image component (DIC). Further, the difference image is encrypted using the result matrices from Lorenz system. The difference image plays an important role in encrypting the original image. Although the difference image does not contain details about the original image, in order to ensure security, the difference image is encrypted and sent to the receiver. These processes are illustrated clearly in Fig. 3.

In the second part of the processes, the DIC is encrypted. The three two-dimensional arrays obtained from Lorenz are XORed with the difference image. However, this step does not encrypt the image; in order to reduce the correlation, we swap the rows and columns horizontally and vertically using Lorenz matrices. Each two-dimensional array, such as Y-Z, X-Y, or X-Z, obtained from Lorenz represents a plane having the same size as the original image components. These arrays are used to shuffle the positions of the pixels for the difference image and lets us call C, the matrix of shuffling positions.

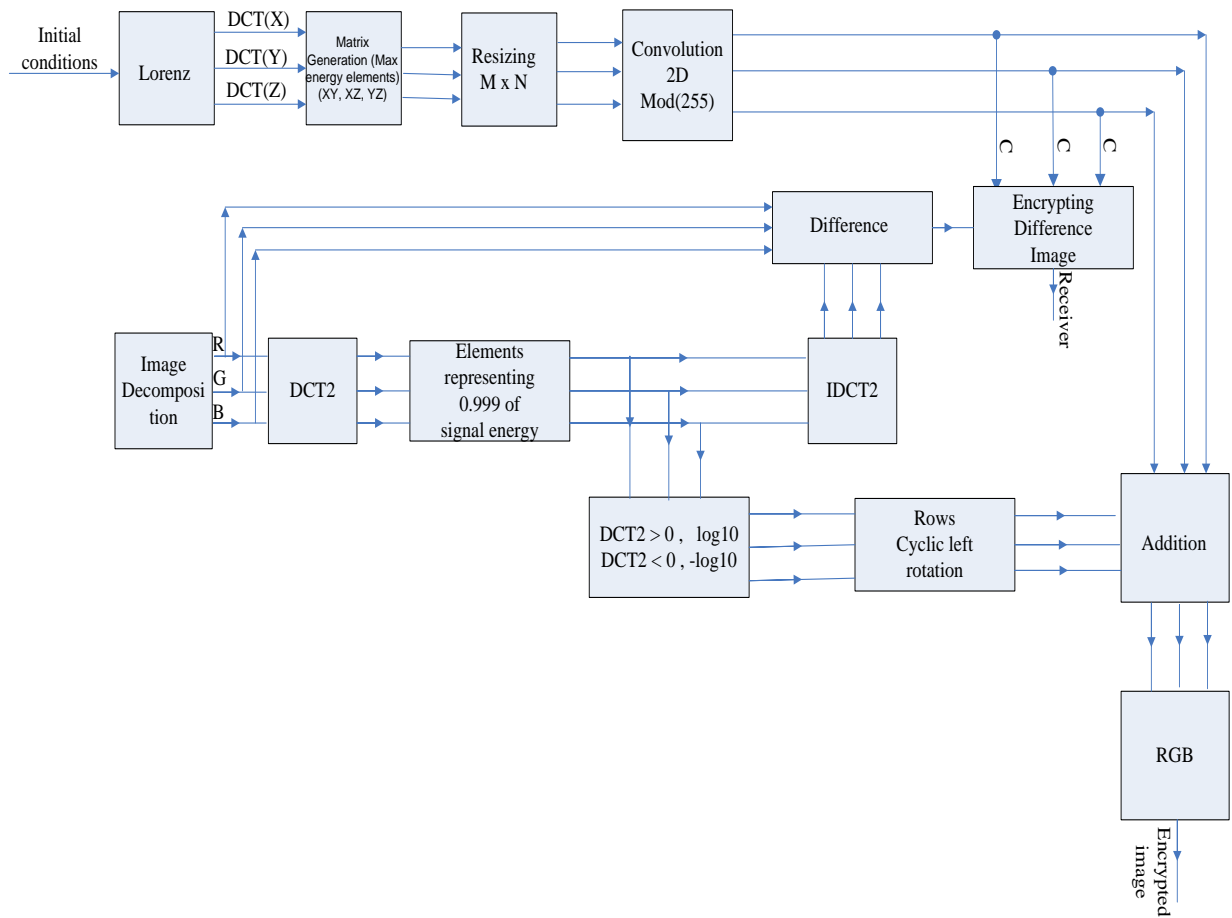

Figure 3. Encryption process

The resulting 2D arrays can be sorted up or down, vertically or horizontally; the shuffling 2D array after being sorted horizontally or vertically can be called A. In this paper, the 2D array is sorted horizontally. The positions of the sorted values in A represent the index for the two-dimensional array of the difference image. The result matrix obtained after performing position-swapping using matrix A can be called B. The original positions of the sorted values in matrix A are used as an index for the image that is obtained from the XOR operation between the difference image and the two-dimensional arrays obtained from Lorenz. In order to perform vertical operations on the matrix, the columns can be transposed horizontally and the same process can be repeated. Thus, the image is shuffled and the correlation between the pixels is reduced. The reduction of the correlation between pixels is important in image encryption; in the simulation results, we provide the correlation factors for the horizontal, vertical, and diagonal directions; further, we compare these correlation factors with the original image correlation factors along these directions and for all colored image components. After the image pixels are obtained from the generated sorted matrix A which was referred to as B, a cyclic left shift is performed between columns or rows, and a predefined number of shifts is performed. Further, the content of the pixels in matrix C is subjected to the same number of predefined cyclic shifts to the left. Then, an XOR operation is performed between the result matrices. This is performed for the three components of the image. The resulting image is called the cipher image. All these processes are illustrated in Figure 4.

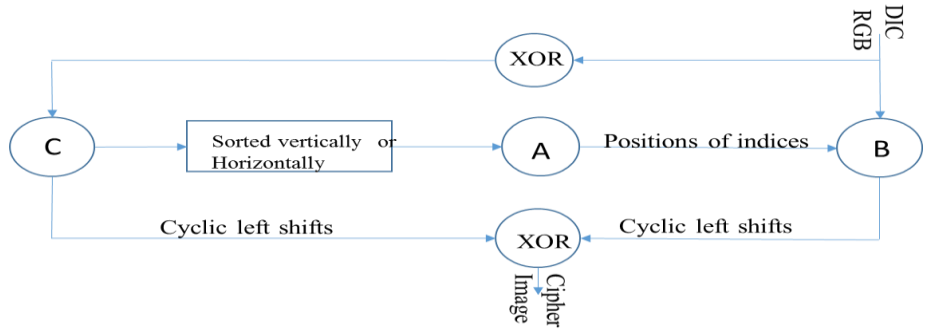

Figure 4. Encryption process for the difference image

Until this step, the difference image is encrypted and the elements that possess high energy in DCT2 are found on the transmitter side. Typically, the values of DCT2 are higher than the pixel values from Lorenz matrices; the DCT2 elements are subjected to a logarithm with base 10, provided that the absolute value is considered for the negative elements and the negative sign is preserved after performing the logarithmic operation$(-log_{10}|M(i,j)|)$. The objective of this step is to reduce the high values of the DCT2 and embed the logarithmic values of DCT2 inside the two-dimensional arrays that are obtained from Lorenz system. This step is represented in (7).

$$\begin{cases} if\ DCT2 > 0,\ log_{10}(M(i,j)) \\ if\ DCT2 < 0, -log_{10}|M(i,j)| \end{cases} \qquad (7)$$

At this stage, the output consists of two-dimensional arrays that are shifted left cyclically at each row from 0 to n-1, where n is equal to the number of columns of the original image. The output of this operation is summed with the output of Lorenz convolved arrays, as shown in Fig. 3. It should be noted that, these processes are repeated three times for the three keys. Then, the composed ciphered image (RGB) is constructed.

## III. DECRYPTION PROCESS

At the receiver, the encrypted constructed image from the highest-energy elements of DCT2 and the encrypted difference image are received. In the decryption process, first, Lorenz system is initialized with the initial conditions using the three keys and the difference image is decomposed into three components. The entire process of decryption is shown in Fig. 5. However, the process involves two steps. These are decryption of DIC and decrypting the image came from the elements of highest energy of DCT2.

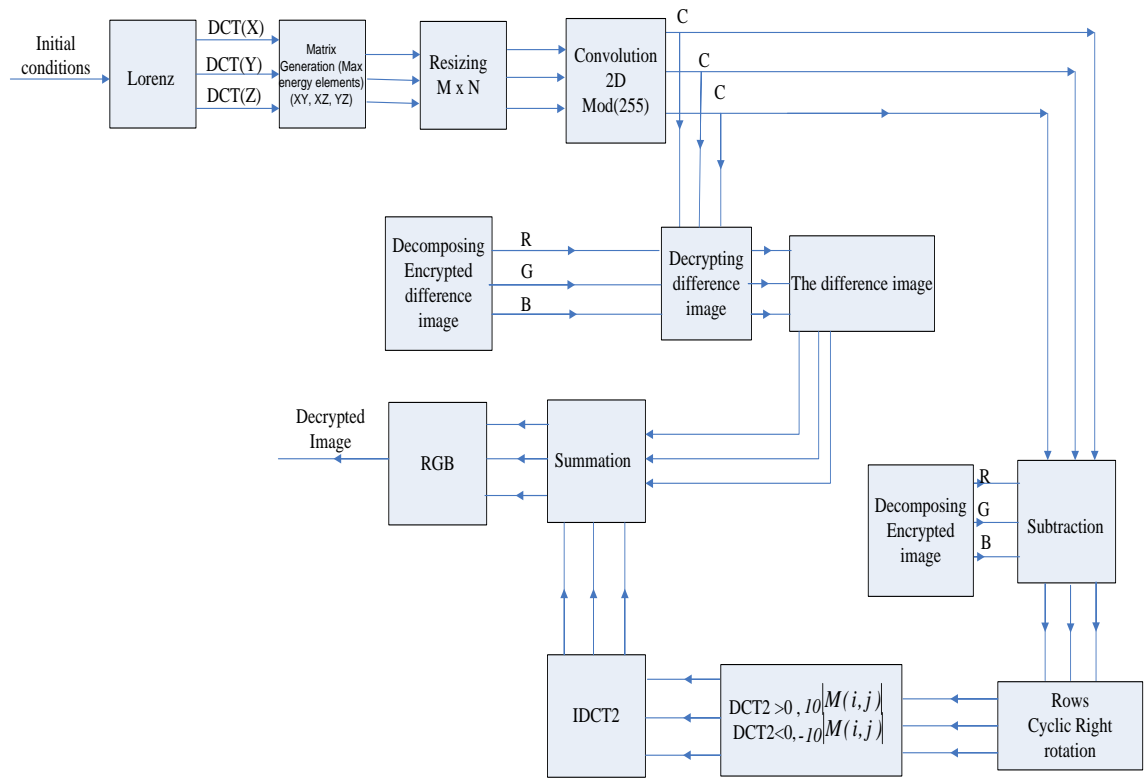

Figure 5. Decryption process

The decryption of the difference image starts with the generation of the position matrices C from the Lorenz system; these secret keys play an important role of matching the position matrices in the entire process. However, in the decryption process, the 2D arrays represented as C in Fig.5 is cyclically shifted right and XORed with the encrypted DIC then the 2D array is cyclically shifted right with the same predefined number of shifts to get B. Here, we have the matrix C and the shuffled B matrix and for each row the positions can be retained. For example, the pixel in first row and in column number 220 came into the position of first pixel during shuffling while in decryption, this pixel is going to column number 220. Also, the output matrix from this operation is called A and will be XORed with C matrix to get the Difference Image Component (DIC).

After the difference image is decrypted, the image came from DCT2 elements is decrypted using arrays in C. The process starts with subtraction of received image from C. This result is subjected to a cyclic shift to the right—as mentioned earlier in the encryption process—from 0 to n-1, where n is the number of columns in the image. The resulting output consists of the high-energy DCT2 elements. The absolute values of these elements are raised to the power 10, while preserving the sign of each element, as shown in (8). Then, inverse of Discrete Cosine Transform (IDCT2) is applied and the resulting image is added to the decrypted difference image.

$$\begin{cases} if\ DCT2 > 0,\ 10^{M(i,j)} \\ if\ DCT2 < 0, -10^{M(i,j)} \end{cases} \quad (8)$$

Where M (i, j) represents the elements of DCT2. IDCT is applied, the difference image and the IDCT2 image are summed, and the RGB decrypted image is displayed, as shown in Fig. 5. It should be noted that these process are applied for each component and repeated three times for the three secret keys.

IV. SIMULATION RESULTS

The Lorenz system described in Section I is initialized using the typical values for the parameters, i.e., ρ=28, σ=10, and β=8/3. Further for simplicity, the initial values of $x_0$, $y_0$, and $z_0$ are truncated for 5 digits and chosen to be 0.84063, 0.13859, and 0.05934, respectively, and the time interval is between 0 and 50, with ε = $10^{-7}$. The Lorenz system has two attractors, as shown in Fig. 1. However, the number of elements having high energy in the DCT for the arrays X, Y, and Z is 208, 228, and 171, respectively. These arrays are combined to yield the new matrices XY, XZ, and YZ. These matrices are resized to match the size of the colored image. Then, the matrices are subjected to interchanging convolution in two dimensions to yield new matrices of the same size as the colored image, i.e., N x N.

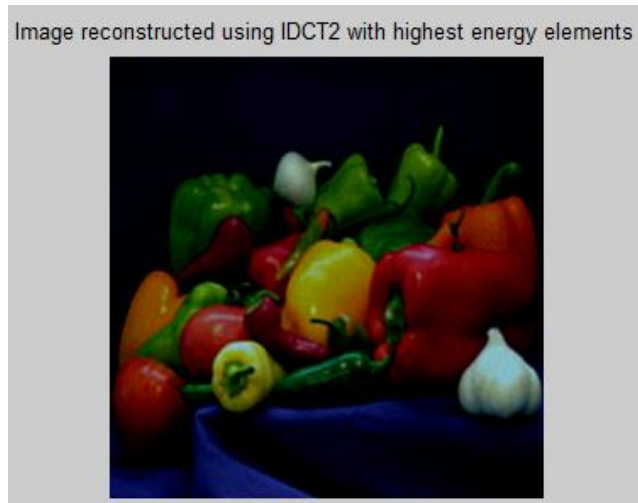

Figure 6. Reconstructed image using IDCT2 with highest-energy elements

The colored image, Peppers, is decomposed into three components, and DCT2 is applied to each component. The maximum-energy elements that constitute 99.9 % of the energy in the image are selected, and the remaining elements are neglected. In a sparse matrix with the highest-energy elements, IDCT is performed; the resulting image is shown in Fig. 6. The size of the original image used for the simulation is 256 x 256; the number of elements used to reconstruct the image in Fig. 5 for the first, second, and third components is 5046, 10635, and 12006, respectively. The difference between the original image and the reconstructed image from IDCT2 is shown in Fig. 7.

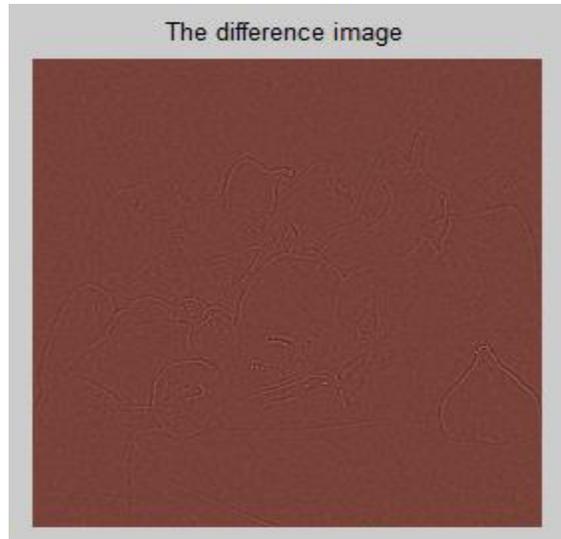

Figure 7. Difference image

The Lorenz matrices are used to encrypt the difference image shown in Fig. 7; for each of the three components, one of the three matrices XY, XZ, or YZ is used to encrypt this component and generate the final encrypted difference image. The encrypted difference image is shown in Fig. 8.

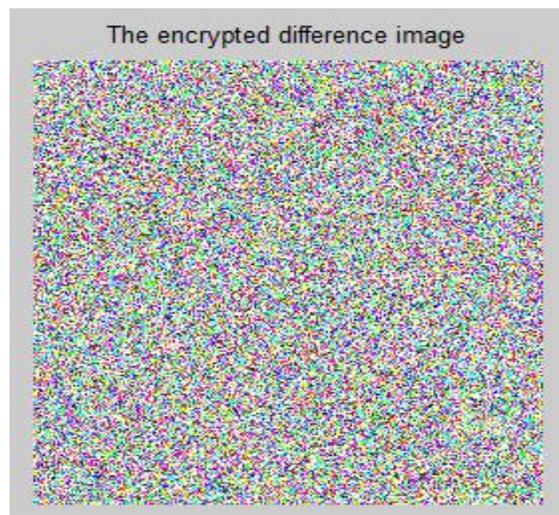

Figure 8. Encrypted difference image

After applying the logarithmic functions in (7) and rotating the rows cyclically from 0 to 255, the DCT2 elements are embedded in the matrix that is obtained from the sum of the difference image and Lorenz matrices.

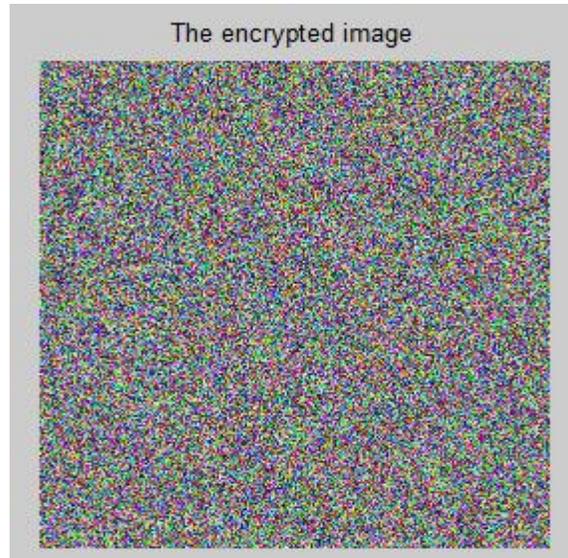
Figure 9. Encrypted image

The encrypted image is shown in Fig. 9. Two encrypted images—the difference image and the resulting image from DCT2—are sent to the receiver. At the receiver, Lorenz initial conditions are used to initialize the Lorenz system and generate the matrices, as explained earlier in the encryption and decryption processes. The decrypted difference image is shown in Fig. 10.

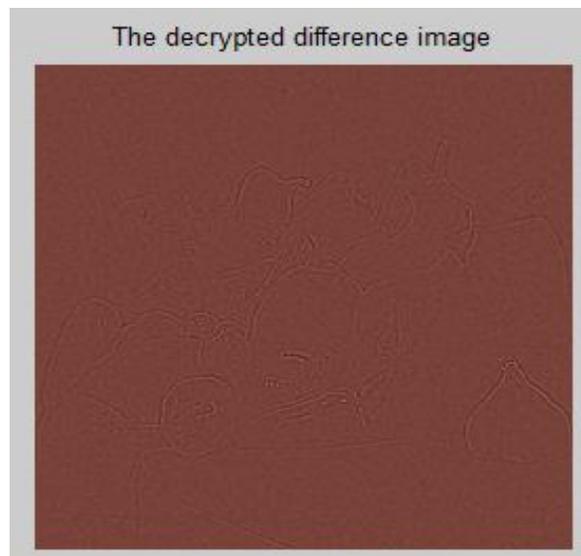
Figure 10. Decrypted difference image

The decomposed encrypted image is subtracted from Lorenz matrices represented as C. Further, a reverse cyclic right shift is performed for each row from 0 to 255, and the anti-logarithmic equations shown in (8) are applied to the resulting image obtained after shifting the rows. Then, IDCT2 is applied to obtain the image shown in Fig. 6. However, the decrypted difference image and the image obtained from IDCT2 are added, and the reconstructed image is shown in Fig. 11. In Fig. 6, the background of the image is not clear and appears dim because some of the energy that has been excluded is preserved there. This observation is clear when Fig. 6 is compared with Fig. 11, which has additional details when the difference image obtained from IDCT2 is added.

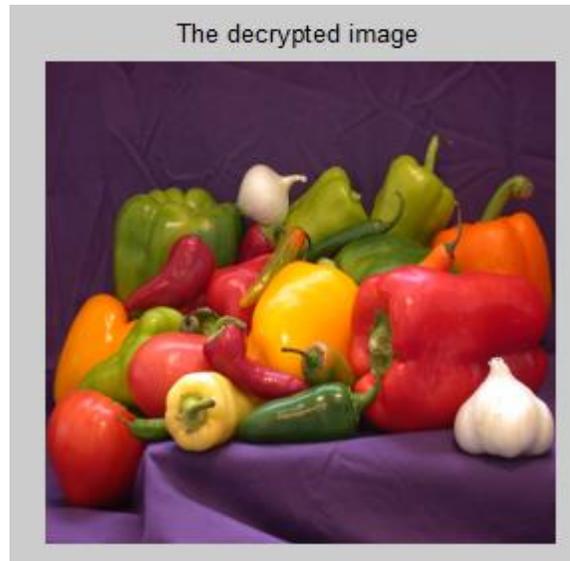

Figure 11. Decrypted final image

## V. STATISTICAL ANALYSIS

In this section, a statistical analysis of the proposed scheme for encryption and decryption of colored images is presented. This section includes histogram analysis, differential attack analysis, and an analysis of the correlation coefficients for the horizontal, vertical, and diagonal directions.

### A. Histogram Analysis

The histogram of an image shows the content and distribution of its pixels. Both histograms of the original and difference images must be compared to ensure that the ciphered image has no clue about its pixel distribution and that any attempt to crack the cipher will fail. Fig. 12 shows the histograms for the difference image and for the ciphered difference image. By comparing the two histograms, we can deduce that the correlation between the ciphered image pixels is almost null, thus hiding all clues related to the statistical distribution of the image. Further, the histograms of the encrypted image and the original Peppers image must be compared, as shown in Fig. 13. From Fig. 12, although a limited band of energy is concentrated between 48 and 75, the pixels in the encrypted image are widely distributed along the higher bands. This distribution is necessary, as mentioned earlier, to hide all clues related to the limited band distribution of the pixels in the difference image.

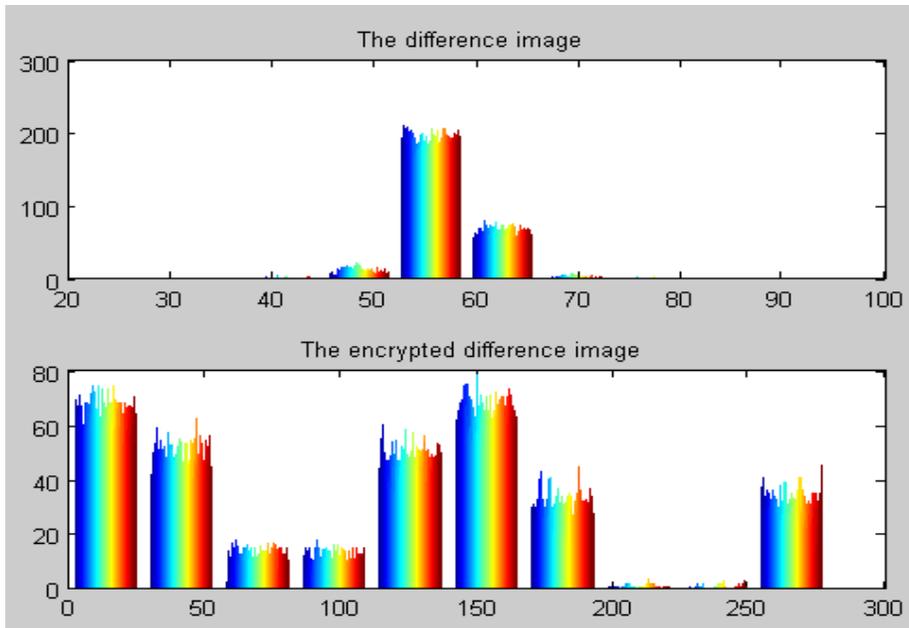

Figure 12. Histograms of the encrypted difference image and the difference image

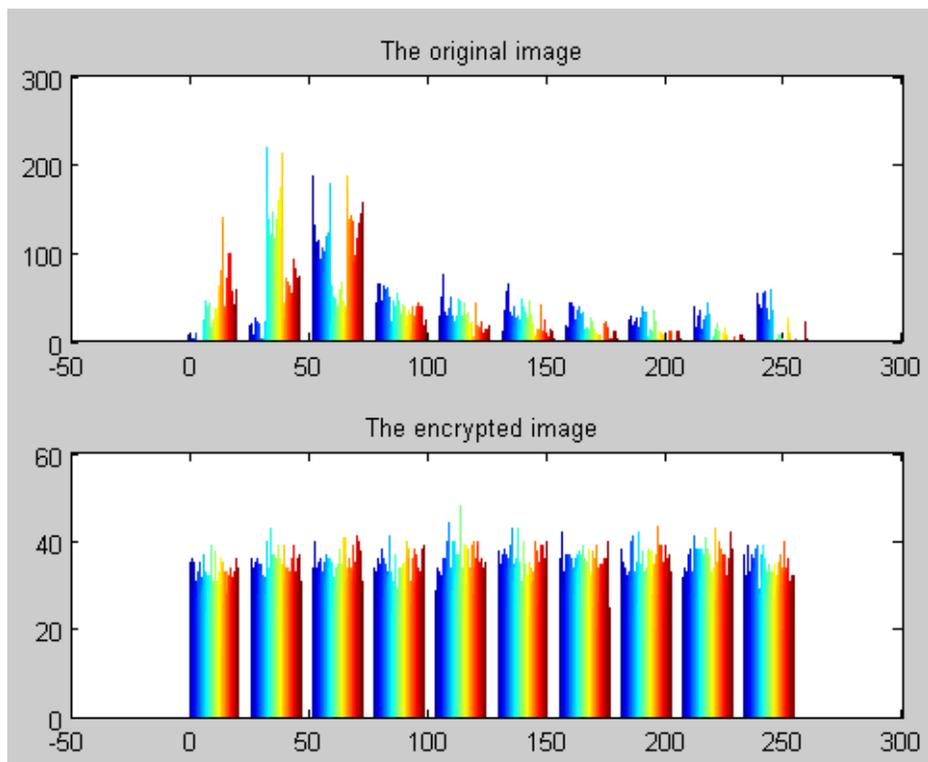

Figure 13. Histograms of the original and the encrypted Peppers image

## B. Correlation Coefficients

Another important factor in the study of the cryptosystem is the correlation factor. The correlation between pixels of the original image is high, whereas the correlation between pixels of the encrypted image should be very low; in the cryptosystem, the pixels should be shuffled well to minimize this correlation. In order to calculate the correlation coefficient for the original image and the encrypted image, the correlation coefficients must be performed between adjacent pixels. Let us suppose that matrices C and D have the same size; then, the correlation coefficient, which is represented as r, can be calculated using (9).

$$r = \frac{\sum_{i=1}^{M}\sum_{j=1}^{N}(C(i,j)-\overline{C})(D(i,j)-\overline{D})}{\left(\left(\sum_{i=1}^{M}\sum_{j=1}^{N}(C(i,j)-\overline{C})^2\right)\left(\sum_{i=1}^{M}\sum_{j=1}^{N}(D(i,j)-\overline{D})^2\right)\right)^{1/2}} \quad (9)$$

Where $\overline{C}$ and $\overline{D}$ are the means of matrix C and matrix D, respectively. The formula in (9) is used to calculate this correlation coefficient horizontally, vertically, and diagonally. The simulation results show a strong correlation coefficient for the original image in the horizontal, vertical, and diagonal directions. The original and encrypted images of the Peppers image are used to calculate the correlation coefficients in the directions mentioned. It is observed that the image has a strong or high correlation factor in all directions, ranging from a maximum of 98.56% along the vertical direction for the first component to a minimum of 91.9% along the diagonal direction for the third component. However, for the encrypted image, the maximum correlation coefficient registered is along the diagonal direction and has a value of 9.19% for the third component; this value is very low when compared with the corresponding value for the original image for the same direction.

TABLE I
CORRELATION COEFFICIENTS FOR THE ENCRYPTED AND ORIGINAL IMAGES FOR FOUR IMAGES

| Colored Image | Metrics | First component | | Second Component | | Third component | |
|---|---|---|---|---|---|---|---|
| | | Original | Encrypted | Original | Encrypted | Original | Encrypted |
| Peppers 256 x 256 | Horz | 0.9854 | 0.0024 | 0.9779 | 0.0033 | 0.9697 | -0.0039 |
| | Vert | 0.9856 | 0.0029 | 0.9791 | -0.0024 | 0.976 | -0.0086 |
| | Diag | 0.9791 | 0.0349 | 0.9783 | -0.0737 | 0.919 | 0.0968 |
| Baboon 256 x 256 | Horz | 0.9171 | 0.0015 | 0.8327 | 0.0087 | 0.9099 | -0.002 |
| | Vert | 0.9472 | -0.0014 | 0.8738 | 0.0008 | 0.9234 | -0.0051 |
| | Diag | 0.8804 | 0.0009 | 0.7982 | 0.0534 | 0.9137 | 0.1278 |
| Circuit 256 x 256 | Horz | 0.5372 | 0.0006 | 0.5324 | 0.00087 | 0.5372 | -0.0029 |
| | Vert | 0.5651 | -0.0066 | 0.5928 | -0.0053 | 0.5757 | -0.0015 |
| | Diag | 0.2312 | -0.0796 | 0.2417 | -0.1413 | 0.2402 | 0.1121 |
| Barbara 256 x 256 | Horz | 0.9641 | -0.0037 | 0.9665 | 0.0081 | 0.9524 | -0.0067 |
| | Vert | 0.9712 | -0.0028 | 0.9702 | -0.005 | 0.9566 | 0.0025 |
| | Diag | 0.9710 | 0.2483 | 0.9484 | -0.1523 | 0.9121 | -0.1238 |

The minimum correlation coefficient is registered along the vertical direction with a value of – 0.86% for the third component, whereas the correlation coefficient has a high value of 97.6% with the same original image and for the same direction. These results are shown in Table I.

Table I provides a comparison of the correlation coefficients for four images—Peppers, Baboon, Circuit Board, and Barbara. The maximum correlation coefficient is 98.56% along the vertical direction for the first component of the original Peppers image, whereas the minimum correlation coefficient is 23.12% along the diagonal direction for the first component of the Circuit Board. For the encrypted images, the maximum correlation coefficient is along the diagonal direction for the first component of Barbara, whereas the minimum correlation coefficient is -15.23% along the diagonal direction for the encrypted second component in Barbara. Circuit Board has dense edges and the correlation coefficients for the original image are low when compared with the correlation coefficients for the other images, i.e., Peppers, Baboon, and Barbara.

The scatter diagram for the first component is shown in Fig. 14. The pixel correlation is high in the original image and very low in the encrypted Peppers image.

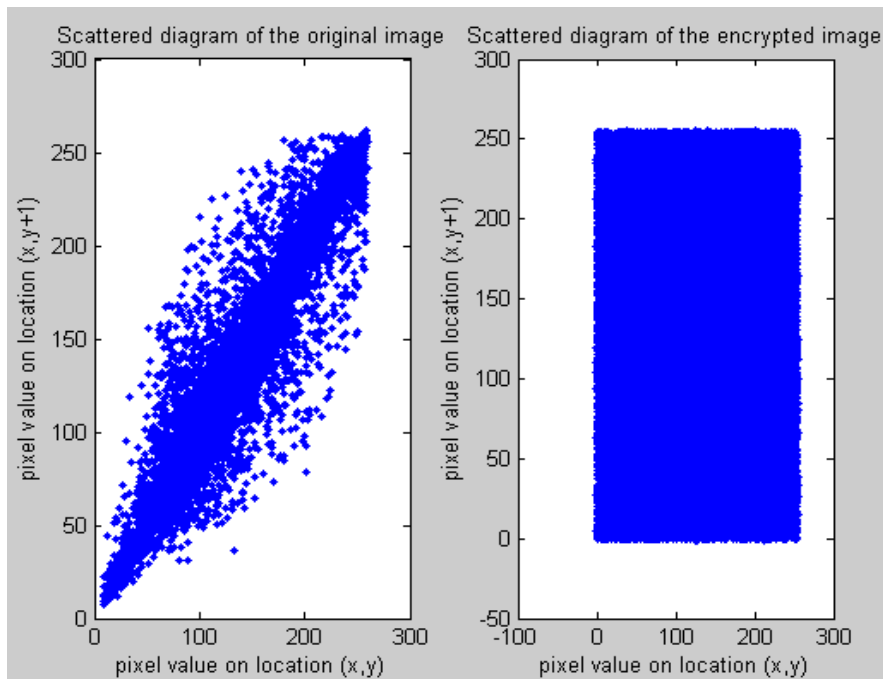

Figure 14. Scatter diagram of the first component of the original and encrypted images

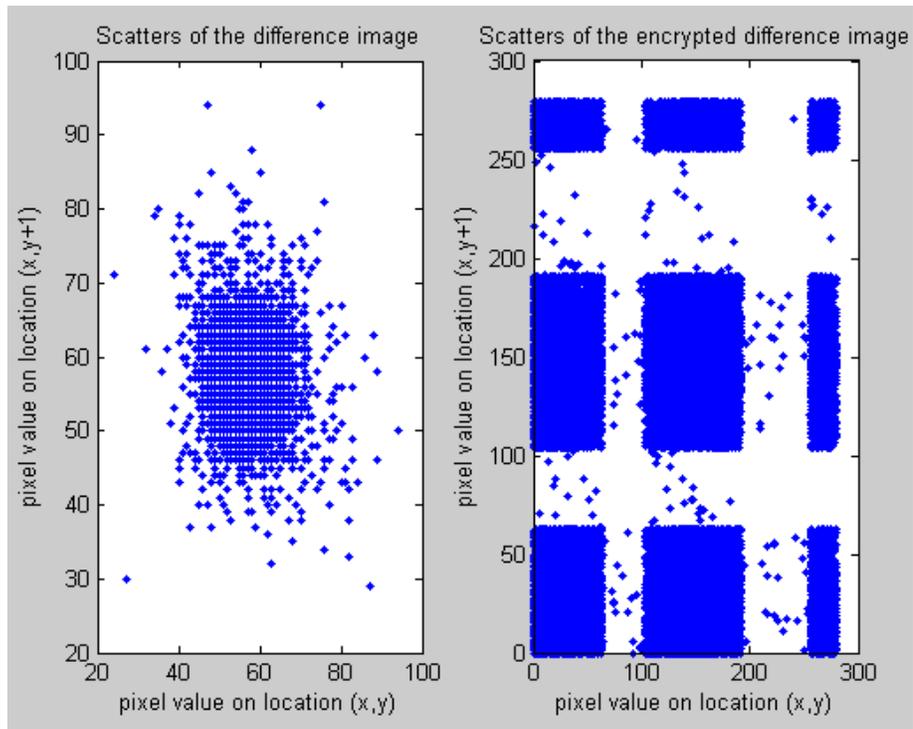

Figure 15. Scatter diagram for the difference and encrypted difference image

For the difference image, the scatter diagram along the horizontal direction for the first component is shown in Fig. 15. From the figure, the pixels in the difference image are compact and concentrated within a limited region, whereas the pixels in the encrypted image are scattered in different regions.

The correlation coefficients for the horizontal, vertical, and diagonal directions obtained from the proposed scheme are compared with the corresponding coefficients obtained from [26] and [27] for the three decomposed components of the images Lena and Temple. The image sizes are the same and are equal to 256 x 256. In Table II, some acronyms—such as Enc. (for encrypted), Orig. (for original), and Prop. (for proposed algorithm)—are used to indicate the type of values in the cells. From the results, for the encrypted images, the proposed scheme shows high potential for decorrelation of the pixels when compared with the other schemes. However, the original correlation coefficients of the original three components are very small when compared with the corresponding coefficients of the encrypted components.

TABLE II
COMPARISON OF CORRELATION COEFFICIENTS FOR THE PROPOSED SCHEME WITH CORRELATION COEFFICIENTS FOR THE SCHEMES IN [26] AND [27]

| Colored Image | Metrics | Type | 1st Compt. | | | 2nd Compt. | | | 3rd Compt. | | |
|---|---|---|---|---|---|---|---|---|---|---|---|
| | | | Prop. | [26] | [27] Avg. | Prop. | [26] | [27] Avg. | Prop. | [26] | [27] Avg. |
| Lena 256 x 256 | Horizontal | Org. | 0.9791 | 0.9572 | 0.9794 | 0.9577 | 0.9432 | 0.9794 | 0.9016 | 0.929 | 0.9794 |
| | | Enc. | -0.0023 | 0.0031 | 92x10-5 | -0.0021 | 0.0011 | 92x10-5 | -0.1344 | 0.0006 | 92x10-5 |
| | Vertical | Org. | 0.9716 | 0.9788 | 0.989 | 0.9431 | 0.9713 | 0.989 | 0.8525 | 0.9559 | 0.989 |
| | | Enc. | -0.0026 | 0.0045 | -0.0021 | 0.0045 | -0.0043 | -0.0021 | -0.007 | 0.0013 | -0.0021 |
| | Diagonal | Org. | 0.9572 | 0.9338 | 0.9694 | 0.9287 | 0.9193 | 0.9694 | 0.784 | 0.9006 | 0.969 |
| | | Enc. | -0.0042 | -0.0054 | 15x10-5 | -0.0054 | -0.0013 | 15x10-5 | -0.102 | -0.002 | 15x10-5 |
| Baboon 256 x 256 | Horizontal | Org. | 0.9649 | 0.9479 | | 0.9335 | 0.9429 | | 0.9085 | 0.9668 | |
| | | Enc. | -0.0061 | 0.0015 | | -0.0058 | -0.0022 | | 0.2111 | 0.0011 | |
| | Vertical | Org. | 0.9621 | 0.9391 | | 0.9283 | 0.9324 | | 0.8925 | 0.9609 | |
| | | Enc. | -0.0005 | 0.0053 | | 0.0073 | -0.0053 | | -0.0104 | -0.0006 | |
| | Diagonal | Org. | 0.9577 | 0.8998 | | 0.9208 | 0.8899 | | 0.8622 | 0.9360 | |
| | | Enc. | 0.0005 | -0.0028 | N/A | -0.0043 | -0.0020 | N/A | -0.0135 | -0.0033 | N/A |

## C. Differential Attack Analysis

Differential attack analysis is one of the methods used to check cipher resistance to attacks. Eli Biham and Edi Shamir [23] laid the foundation for these attacks on block ciphers. The cipher resistance to differential attacks is measured based on two factors. The first factor is called the number of pixels change rate (NPCR), and the second factor is called the unified average changing intensity (UACI). If we have two images of size M x N and either plane or ciphered images represented as $C_1$ and $C_2$, and we would like to define a bipolar array represented as D and having the same matrix size as the plane or ciphered image, then D has a value of either 0 or 1. In order to determine whether the value of D is 0 or 1, a condition is applied. If $C_1(i, j) = C_2(i, j)$, D =0; otherwise, D=1. As implied by its name, NPCR is calculated by counting the number of pixels that have changed in both the images and dividing this value by M x N. Thus, NPCR can be determined by using (10).

$$NPCR = \frac{\sum_{i=1}^{M}\sum_{j=1}^{N} D(i,j)}{M \times N} \times 100 \% \qquad (10)$$

UACI can be represented by the following equation:

$$UACI = \frac{1}{M \times N} \left[\sum_{i=1}^{M}\sum_{j=1}^{N} \frac{|c_1(i,j) - c_2(i,j)|}{255}\right] \times 100 \% \qquad (11)$$

where $c_1(i, j)$ and $c_2(i, j)$ are the first and second cipher (or plane) images, respectively. The mean absolute error can be represented by (12).

$$MAE = \frac{1}{M \times N} \sum_{i=1}^{M}\sum_{j=1}^{N} |C_1(i,j) - C_2(i,j)| \qquad (12)$$

TABLE III
CALCULATION OF NPCR, UACI, AND MAE FOR FOUR IMAGES, INCLUDING PEPPER

| Colored Image 256 x 256 | Metrics | 1st Component | 2nd Component | 3rd Component |
|---|---|---|---|---|
| Peppers | UPCR | 100 | 100 | 99.9954 |
| | UACI | 31.7526 | 34.8253 | 35.1903 |
| | MAE | 80.9692 | 88.8045 | 89.7353 |
| Baboon | UPCR | 100 | 100 | 100 |
| | UACI | 29.4949 | 27.8191 | 30.529 |
| | MAE | 75.2119 | 70.9386 | 77.8489 |
| Circuit | UPCR | 100 | 100 | 100 |
| | UACI | 33.693 | 30.8795 | 31.9975 |
| | MAE | 85.9172 | 78.7428 | 81.5936 |
| Barbara | UPCR | 99.6628 | 99.7025 | 99.765 |
| | UACI | 31.8762 | 36.3306 | 37.3611 |
| | MAE | 81.2842 | 92.643 | 95.2709 |

NPCR, UACI, and mean absolute error (MAE) were calculated for two images—Pout and Cameraman. The results are shown in Table III. Table III also shows the values of NPCR, UACI, and MAE for the colored images of Baboon, Circuit Board, and Barbara.

The differential attack analysis of the three decomposed components of the Lena and Temple images is performed; the results of the analysis for the proposed scheme are compared with the corresponding results for the schemes in [26] and [27] in terms of UPCR, UACI, and MAE. The image sizes are the same and are equal to 256 x 256. The results are shown in Table IV. Some values are not available in the references; they are indicated as not available (N/A).

TABLE IV
COMPARISON OF DIFFERENTIAL ATTACK ANALYSIS OF THE PROPOSED SCHEME WITH SCHEMES IN [26] AND [27]

| Colored Image | Metrics | 1st Compt. | | | 2nd Compt. | | | 3rd Compt. | | |
|---|---|---|---|---|---|---|---|---|---|---|
| | | Prop. | [26] | [27] Avg. | Prop. | [26] | [27] Avg. | Prop. | [26] | [27] Avg. |
| Lena 256 x 256 | UPCR | 100 | 99.557 | 99.598 | 100 | 99.517 | 99.598 | 84.105 | 99.513 | 99.598 |
| | UCAI | 32.982 | 33.557 | 33.412 | 30.261 | 33.453 | 33.412 | 77.166 | 33.388 | 33.412 |
| | MAE | 84.105 | N/A | N/A | 27.418 | N/A | N/A | 69.916 | N/A | N/A |
| Temple 256 x 256 | UPCR | 100 | 99.504 | N/A | 100 | 99.574 | N/A | 100 | 99.494 | N/A |
| | UCAI | 34.361 | 33.539 | | 32.650 | 33.488 | | 31.461 | 33.351 | |
| | MAE | 87.621 | N/A | | 83.258 | N/A | | 80.226 | N/A | |

*D. Information Entropy*

The information content of a signal is called the entropy. It describes the information redundancy associated with feature randomness. For example, an entropy of 7.35 bits/pixel for an image describes the statistical average number of bits required to represent the pixels in the image; this value is 7.35 bits. The entropy can be represented mathematically as follows [24]:

$$H(s) = \sum_{i=0}^{2^N-1} p(s_i) \log_2(\frac{1}{p(s_i)}) = -\sum_{i=0}^{2^N-1} p(s_i) \log_2(p(s_i)) \qquad (13)$$

Where $p(s_i)$ is the occurrence probability of a pixel in an image, N is the length of a pixel in binary numbers (typically, N=8 for a gray-scale image), and H(s) is the entropy described in bits per pixel or symbol. An essential property of a cryptosystem is its ability to resist entropy attack; the ideal value of entropy of the encrypted images is 8 bits/pixel [24].

TABLE V
ENTROPY OF FOUR COLORED IMAGES, INCLUDING PEPPERS, AND THEIR CORRESPONDING ENCRYPTED IMAGES

| Colored Image 256 x 256 | Entropy | | | | |
|---|---|---|---|---|---|
| | Metrics | 1st Component | 2nd Component | 3rd Component | Average |
| Peppers | Original | 7.1785 | 7.0133 | 6.7975 | 6.7975 |
| | Encrypted | 7.9928 | 7.9939 | 7.9935 | 7.9952 |
| Baboon | Original | 7.6058 | 7.358 | 7.6662 | 7.5433 |
| | Encrypted | 7.9942 | 7.9943 | 7.9947 | 7.9944 |
| Circuit Board | Original | 7.6759 | 7.6077 | 7.5505 | 7.6114 |
| | Encrypted | 7.9949 | 7.9952 | 7.9957 | 7.9953 |
| Barbara | Original | 6.42 | 6.4457 | 6.3807 | 6.4155 |
| | Encrypted | 7.9936 | 7.9935 | 7.9949 | 7.994 |

TABLE VI
COMPARISON OF ENTROPY OF THE PROPOSED SCHEME WITH THE SCHEMES IN [26] AND [27]

| Colored Image | Type | Entropy | | | | | | | | |
|---|---|---|---|---|---|---|---|---|---|---|
| | | 1st Compt. | | | 2nd Compt. | | | 3rd Compt. | | |
| | | Prop. | [26] Avg. | [27] Avg. | Prop. | [26] Avg. | [27] Avg. | Prop. | [26] Avg. | [27] Avg. |
| Lena 256 x 256 | Org. | 7.241 | 7.2417 | N/A | 7.5766 | 7.2417 | N/A | 6.9167 | 7.2417 | N/A |
| | Enc. | 7.994 | 7.9971 | 7.9998 | 7.9938 | 7.9971 | 7.9998 | 7.9938 | 7.9971 | 7.9998 |
| Temple 256 x 256 | Org. | 7.795 | 7.4032 | | 7.9028 | 7.4032 | | 7.9156 | 7.4032 | |
| | Enc. | 7.994 | 7.9971 | N/A | 7.9932 | 7.9971 | N/A | 7.9935 | 7.9971 | N/A |

Table V shows the entropy of some chosen colored images—Peppers, Baboon, Circuit Board, and Barbara—that have the same size of 256 x 256 pixels and their corresponding encrypted images for the three components (red, green, and blue) of the colored images. The entropy of the ciphered images is close to 8 bits/pixel. Thus, the cryptosystem can resist an information entropy attack.

Further, an entropy comparison between the proposed scheme and the schemes in [26] and [27] is performed for the original and the encrypted images of Lena and Temple. The results are shown in Table VI. All the images have a size of 256 x 256. Some values are not available and are indicated as N/A. However, for the schemes in [26] and [27], the average values of the entropies are calculated and are shown in the table as the average for all decomposed components. The values show that the average number of bits used to represent the encrypted images is approximately 8 bits/pixel.

E. *Peak Signal-to-Noise Ratio (PSNR)*

The peak signal-to-noise ratio (PSNR) is defined as the ratio between the maximum value of the signal and the root mean squared error expressed in decibels (dB). PSNR can be determined by the following equation [25]:

$$PSNR = 20\ log10\left(\frac{MAX_{f(x,y)}}{\sqrt{MSE}}\right) \quad (14)$$

where $MAX_{f(x,y)}$ is the maximum value of the original image f(x, y), and MSE is the mean squared error, which can be expressed by the following equation [25]:

$$MSR = \frac{1}{m.n}\sum_{i=1}^{m}\sum_{j=1}^{n}|f(i,j) - g(i,j)|^2 \quad (15)$$

where f(i, j) represents the original image, and g(i, j) represents the decrypted or recovered image. The PSNRs for different colored images are calculated for the encrypted and decrypted images and for the three components—red, green, and blue. The images used are Peppers, Circuit Board, Baboon, and Barbara. The results are shown in Table VII. The MSE between the components of the original images and the components of the encrypted or decrypted images is calculated based on (15); then, (14) is applied. In the table, ENC and DEC denote encrypted and decrypted, respectively.

TABLE VII
COMPARISON OF PSNRs OF SOME ENCRYPTED AND DECRYPTED COLORED IMAGES, INCLUDING PEPPERS

| Colored Image 256 x 256 | PSNR (dB) | | | | |
|---|---|---|---|---|---|
| | Metrics | 1st Component | 2nd Component | 3rd Component | Average |
| Peppers | ENC | 5.5749 | 9.794 | 11.2443 | 8.8713 |
| | DEC | 59.174 | 59.0424 | 59.0392 | 59.0852 |
| Baboon | ENC | 4.7628 | 4.1589 | 5.9392 | 4.954 |
| | DEC | 58.9141 | 57.6225 | 58.8304 | 58.4556 |
| Circuit Board | ENC | 8.8595 | 7.9615 | 8.6688 | 8.4966 |
| | DEC | 60.0569 | 59.8431 | 59.9435 | 59.9479 |
| Barbara | ENC | 9.2677 | 10.9145 | 11.3207 | 10.501 |
| | DEC | 58.8955 | 58.2787 | 57.6959 | 58.29 |

The minimum PSNR among the four encrypted images is obtained for the first component of Baboon and has a value of 4.7628 dB and an average of 4.954 dB for all the three components, whereas the maximum PSNR value is obtained for Barbara and has a value of 10.501 dB.

VI. CONCLUSIONS

During the past decade, chaos has attracted a tremendous amount of research attention. Owing to the deterministic nature of chaos, it can be used for encryption. The Lorenz system is a chaotic system that produces chaos. In this study, the Lorenz system is biased using initial conditions which came from three different keys and produces matrices that are used to encrypt and decrypt the colored difference image. This difference image is obtained by decomposing the colored image into three components, applying DCT2 to each component, obtaining the elements with high energy, and applying IDCT2. The difference between the original image components and the recovered image from IDCT2 is computed. This result is encrypted and sent to the receiver. The DCT2 elements that have high energy are subjected to the logarithmic function, cyclically shifted horizontally, and embedded in the sum of the difference image and Lorenz matrices. During the decryption process, the difference image is decrypted and summed with the Lorenz matrices; then, it is subtracted from the encrypted image to regenerate the DCT2 elements. These elements are cyclically shifted in reverse—i.e., to the right along the horizontal direction—and are raised to the power 10 while preserving the values of the DCT2 coefficients. Then, IDCT2 is applied and the image is summed with the difference image to produce the final decrypted image. Statistical results show that the correlation between pixels is greatly reduced and yields negative values for some components.

REFERENCES

[1] Q. Ding, Q. Zhang, and W. Li, "Discrete design of Lorenz chaotic system based on Euler method and image encryption," *3rd Int. Conf. on Robot, Vision and Signal Processing*, IEEE, 978-1-4673-9646-2/15, doi: 10.1109/RVSP.2015.4.
[2] J. Ahmad, M. A. Khan, S. O. Hwang, and J. S. Khan, "A compression sensing and noise-tolerant image encryption scheme based on chaotic maps and orthogonal matrices," *Neural Comput & Applic.*, doi: 10.1007/s00521-016-2405-6, 2016.
[3] S. B. Matondo, "Two-level chaos-based cryptography for image security," Master's Degree Thesis, 2012, F'SATI, Tshwane University of Technology.
[4] A. Akhavan, A. Samsudin, and A. Akhshani, "Cryptanalysis of "an improvement over an image encryption method based on total shuffling"," *Optics Communications,* 350, pp. 77–82, Sep. 2015.
[5] M. A. Murillo-Escobar, C. Cruz-Hernández, F. Abundiz-Pérez, R. M. López-Gutiérrez, and O. R. Acosta Del Campo, "A RGB image encryption algorithm based on total plain image characteristics and chaos," *Signal Processing*, 109, pp. 119–131, Apr. 2015.
[6] S. Som, S. Dutta, R. Singha, A. Kotal, and S. Palit, "Confusion and diffusion of color images with multiple chaotic maps and chaos-based pseudorandom binary number generator," *Nonlinear Dynamics*, 80, pp. 615–627, Apr. 2015.
[7] W. Yao, X. Zhang, Z. Zheng, and W. Qiu, "A colour image encryption algorithm using 4-pixel Feistel structure and multiple chaotic systems," *Nonlinear Dynamics*, Mar. 2015.
[8] W. Zhang, H. Yu, and Z. L. Zhu, (2015) "Color image encryption based on paired interpermuting planes," *Optics Communications*, 338, pp. 199–208, Mar. 2015.
[9] T. Xiang, J. Hu, and J. Sun, "Outsourcing chaotic selective image encryption to the cloud with steganography," *Digital Signal Processing*, 43, pp. 28–37, Aug. 2015.
[10] Y. Liang, G. Liu, N. Zhou, and J. Wu, "Image encryption combining multiple generating sequences controlled fractional DCT with dependent scrambling and diffusion," *Journal of Modern Optics*, 62, pp. 251–264, Feb. 2015.
[11] X. Zhang, X. Fan, J. Wang, and Z. Zhao, "A chaos-based image encryption scheme using 2D rectangular transform and dependent substitution," *Multimedia Tools and Applications*, Nov. 2014.
[12] W. Zhen, H. Xia, L. Yu-Xia, and S. Xiao-Na, "A new image encryption algorithm based on the fractional-order hyperchaotic Lorenz system," *Chinese Physics B*, vol. 22, no.1, 010504 (2013), doi:10.1088/1674-1056/22/1/010504.
[13] C. K. Huang and H. H. Nien, "Multi chaotic systems based pixel shuffle for image encryption," *Optics Communications*, vol. 282, no. 11, 1 June 2009, pp. 2123–2127.
[14] D. Li, J. A. Lu, X. Wu, and G. Chen, "Estimating the bounds for the Lorenz family of chaotic systems," *Chaos, Solitons & Fractals*, vol. 23, no. 2, Jan. 2005, pp. 529–534.
[15] Z. H. Guan, F. Huang, and W. Guan, "Chaos-based image encryption algorithm," *Physics Letters A*, vol. 346, no. 1–3, 10 October 2005, pp. 153–157.
[16] I. Grigorenko and E. Grigorenko, "Chaotic Dynamics of the Fractional Lorenz System," *Phys. Rev. Lett.*, 91, 034101 – Published 17 July 2003.

[17] S. K. Yang, C. L. Chen, and H. T. Yau, "Control of chaos in Lorenz system," *Chaos, Solitons & Fractals*, vol. 13, no. 4, March 2002, pp.767–780.

[18] D. Sundararajan, *The Discrete Fourier Transform*. World Scientific Publi. Singapore, 2001, pp. 303–311, doi: 10.1142/9789812810298_0015.

[19] J. Fridrich, "Symmetric ciphers based on two-dimensional chaotic maps," *Int. J. Bifurcation Chaos*, 08, 1259, 1998, doi: 10.1142/S021812749800098X.

[20] Y. Zheng and S. H. Singh, "Adaptive control of chaos in Lorenz system," *Dynamics and Control*, 7, pp. 143–154, 1997.

[21] K. M. Cuomo and A. V. Oppenheim, "Circuit implementation of synchronized chaos with applications to communications," *Phys. Rev. Lett.*, 71, 65 – Published 5 July 1993.

[22] T. Yang, L. B. Yang, and C. M. Yang, "Impulsive control of Lorenz system," *Physica D: Nonlinear Phenomena*, vol. 110, no. 1–2, 1 December 1997, pp. 18–24.

[23] E. Biham, and A. Shamir "Differential cryptanalysis of the data encryption standard", Springer-Verlag, 1993, Vol. 28. New York

[24] N. Zhou, A. Zhang, F. Zheng, and L. Gong, "Novel image compression-encryption hybrid algorithm based on key-controlled measurement matrix in compressive sensing," *Optics & Laser Technology*, 2014, 62, pp. 152–160.

[25] H. Huang, G. Ye, H. Chai, and O. Xie, "Compression and encryption for remote sensing image using chaotic system," *Security and Communication Networks*, 2015, vol. 8, 3659–3666, doi: 1002/sec. 1289.

[26] A. M. Awad, R. F. Hassan, and A. M. Sagheer, "New Image Encryption Technique Based on Wavelet / DCT Transforms Using Lorenz Chaotic Map (NIETWDL)," *Int. Jour. of Computer Science and Information Security (IJCSIS)*, June 2016, vol. 14, no. 6.

[27] S. Karpate and A. Barve, "A novel encryption algorithm using chaotic Lorenz attractor and Knights tour," *ICCCT*, 2015, ACM 978-1-4503-3552-2/15/09.